\documentclass[fleqn,12pt,twoside]{article}
\usepackage[headings]{espcrc1}

\readRCS
$Id: espcrc1.tex,v 1.2 2004/02/24 11:22:11 spepping Exp $
\usepackage{graphicx}
\usepackage[figuresright]{rotating}

\newcommand{\AmS}{{\protect\the\textfont2
  A\kern-.1667em\lower.5ex\hbox{M}\kern-.125emS}}
\def\be{\begin{eqnarray}}
\def\ee{\end{eqnarray}}
\def\lsim{\mathrel{\rlap{\lower3pt\hbox{\hskip1pt$\sim$}}
     \raise1pt\hbox{$<$}}} 
\def\gsim{\mathrel{\rlap{\lower3pt\hbox{\hskip1pt$\sim$}}
     \raise1pt\hbox{$>$}}}
\title{The Problem of Mass: Mesonic Bound States Above $T_c$\thanks{We are
grateful to Bielefeld lattice group, especially to Felix Zantow, for the lattice gauge results.}}

\author{H.-J. Park\address[pnu]{Department of Physics,
        Pusan National University, Busan 609-735, Korea},
        C.-H. Lee\addressmark\thanks{HJP and CHL were supported by
          grant No. R01-2005-000-10334-0(2005) from the Basic Research
          Program of the Korea Science \& Engineering Foundation.},
        G.E. Brown\address[suny]{Department of Physics and Astronomy,
        SUNY, Stony Brook, NY11794, USA}\thanks{GEB was
        supported in part by the US Department of Energy
         under Grant No.DE-FG02-88ER40388. },
         and
        M. Rho\address{Service de Physique Th\'eorique,
        CEA Saclay, 91191 Gif-sur-Yvette c\'edex, France}
        }

\runtitle{The Problem of Mass: Mesonic Bound States Above $T_c$}
\runauthor{Park, Lee, Brown \& Rho}

\begin{document}

\maketitle

\begin{abstract}
By extending the Bielefeld LGS (Lattice Gauge Simulation) color
singlet interaction, we find that the masses of $\pi, \sigma,
\rho$ and $a_1$ excitations, 32 degrees of freedom in all, go to
zero (in the chiral limit) as $T\rightarrow T_c$. This result
indicates a smooth phase transition at $T_c$, at which from above
the masses and couplings of mesons vanish \`a la Brown-Rho
scaling. We discuss that our scenario successfully explains the
STAR (STAR Collaboration) $\rho^0/\pi^-$ ratio in Au-Au peripheral
collisions at RHIC.
\end{abstract}

\section{Introduction}

The problem of mass is one of the most fundamental in physics.
We now have experimental evidence that meson masses decrease by
$\sim 20 \%$ as the density increases to nuclear matter density. This
incipient decrease has been seen in the STAR
data for the $\rho$-meson,
at a low density $\sim 0.15 n_0$, where $n_0$ is nuclear
matter density \cite{Shuryak2003}.
Data on in-medium $\omega$ photoproduction in $Nb$ measured by
the CBELSA/TAPS collaboration  shows this mass to decrease roughly
consistently with about $15\%$ for nuclear matter density
\cite{Trnka-Metag}.
Such decrease in mass was predicted by Brown and Rho \cite{BR91}.
This is well and fine for finite density, although the experiments take us up
to only $\sim n_0$. However, Harada and Yamawaki \cite{HY:PR}
carrying out a renormalization group calculation in their vector
manifestation show that the vector mass goes to zero at a fixed
point as the temperature goes up to $T_c$ from below.

Now the question is what happens when one approaches $T_c$ from
above. The existence of mesonic bound states has been predicted by
many authors \cite{Kunihiro,asakawaetal,BLRS}. In this work, we
discuss the masses of color-singlet mesonic bound states above
$T_c$ \cite{PLB} and their implications on STAR at RHIC
\cite{BLR:STAR}.

\section{Masses of Mesonic Bound States Above $T_c$}

\begin{figure}[tb]
\begin{minipage}[t]{82mm}
\includegraphics[height=2.5in]{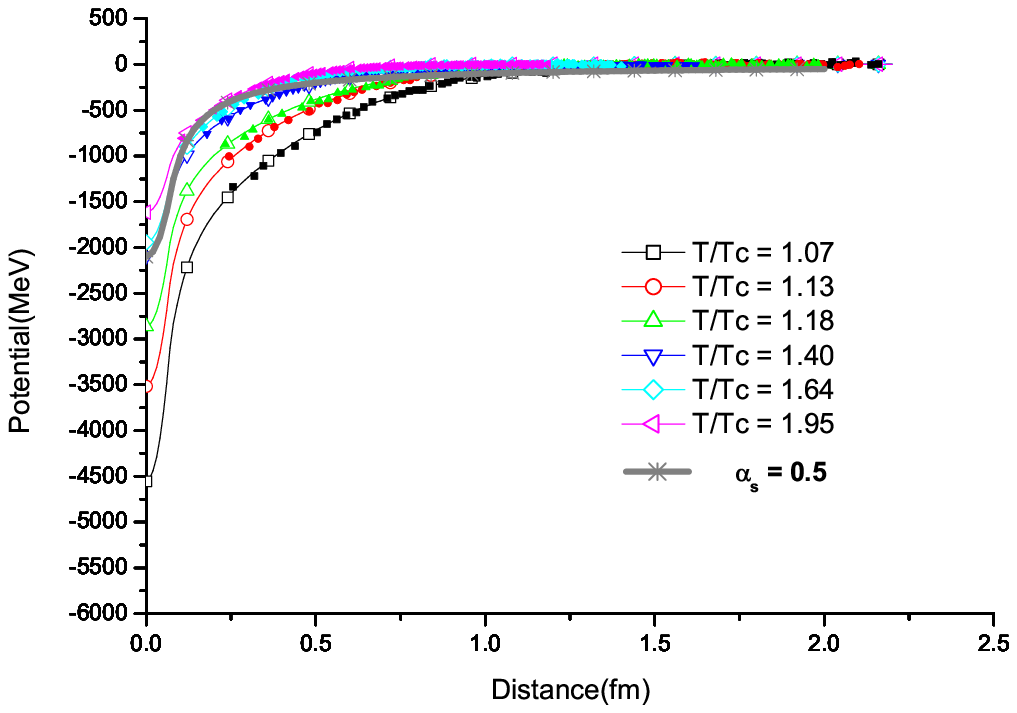}
\caption{The potential extracted from Bielefeld lattice gauge
simulation \cite{Kaczmarek2004b} (filled symbols)
with $m_q =1.4$ GeV.
Thin solid lines are fitted curves and
the thick solid line is for the color-Coulomb interaction
with constant coupling $\alpha_s=0.5$ \cite{BLRS}.}
\label{fig1}
\end{minipage}
\hspace{\fill}
\begin{minipage}[t]{73mm}
\includegraphics[height=2.5in]{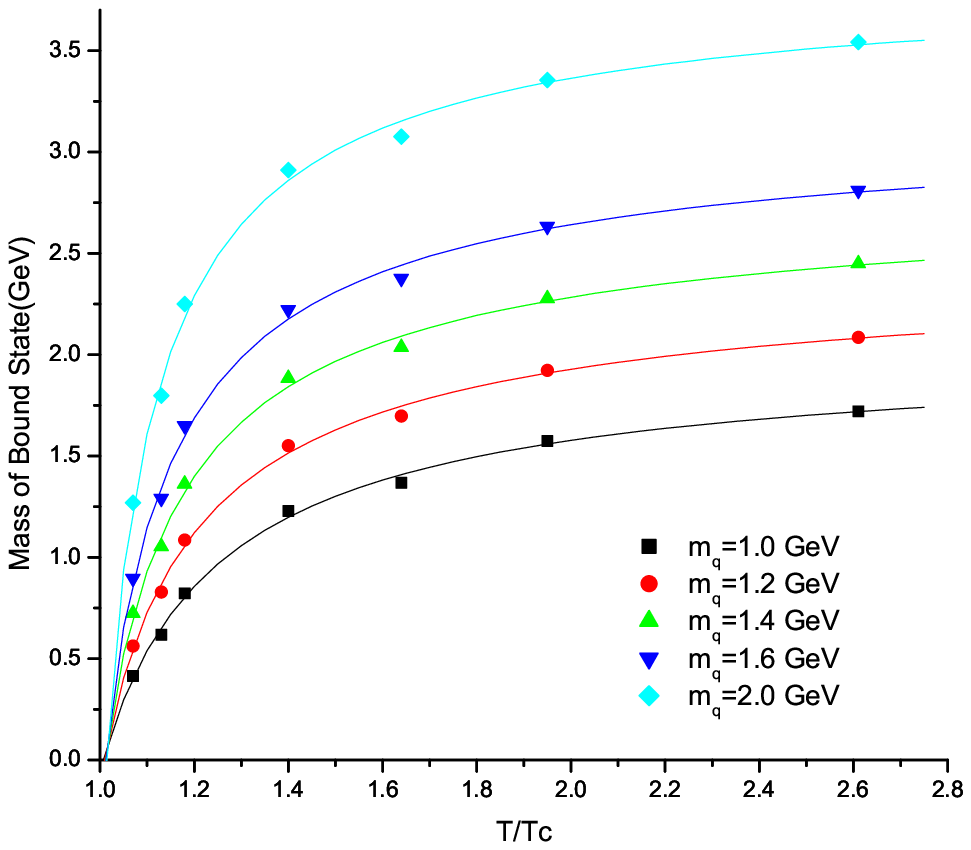}
\caption{Mass of bound states of chirally restored SU(2) quarks
and antiquarks. The fitting curves show that the mass of bound
states approaches zero as $T$ goes to $T_c$, independently of the
thermal quark masses.} \label{fig2}
\end{minipage}
\end{figure}

The Bielefeld group \cite{KKZP}
have carried out lattice
gauge calculations to obtain the heavy quark free energy
for the region of temperatures above $T_c$.
We analyzed their results for the color singlet
(Coulomb) potentials \cite{PLB}.
The finite temperature analog of the static potential is not known and
it is not clear whether at all it can be properly defined. Therefore
we use the internal energy as potential in the Klein-Gordon equation.
The color singlet internal energy can be derived from the free energy
obtained in lattice gauge calculations:
\be
V_{1}(r,T)=F_1 (r,T)- T \frac{\partial F_1 (r,T)}{\partial T}
\label{V1}
\ee
Since we fitted the Bielefeld LGS data, final results are not sensitive
to the choice of different parameterizations. 
In the above expression (Eq.~(\ref{V1})) we subtracted the value of the
free energy at $r\rightarrow \infty$, namely $F_1 (r=\infty,T)$.
This is because we are only interested in the binding energy.
The resulting potential is summarized in Fig.~\ref{fig1}.
As in Brown et al.\cite{BLRS}, in order to enforce the asymptotic
freedom at the origin, we used the molecular radius as $R \simeq
{\hbar}/{2m_q}$. Inside the molecular radius we introduced the
effective potential which is the same as the Coulomb potential for
a uniform charge distribution \cite{BLRS}.

Lattice calculations from the Bielefeld group \cite{KKZP} are the
interactions between heavy quarks, so we need to add (model
dependent) interactions so as to use them for the light quarks
making up our mesons. Brown \cite{Brown52} showed that in a
stationary state the E.M. interaction Hamiltonian between fermions
is \be H_{int} = \frac{e^2}{r} ( 1- \vec{\alpha}_{1} \cdot
\vec{\alpha}_{2}), \label{Hint} \ee where $\vec{\alpha}_{1,2}$ are
velocity operators. This formula, suitably transcribed to QCD, is
applicable to the system we are considering. In this paper we
restrict our consideration to $T\sim T_c$. It will turn out that
the mesons are essentially massless in this region. Since $\vec
\alpha$ is essentially the helicity, we expect for the light quark
potential to be \be V_{\rm light}^{\rm eff} (T=T_c) &=& \left\{
\begin{array}{lll} 2 V_{\rm heavy}^{\rm eff} & \qquad for &
\vec{\alpha}_{1} \cdot \vec{\alpha}_{2}=-1
\\ 0 &\qquad  for& \vec{\alpha}_{1} \cdot \vec{\alpha}_{2}=+1
\end{array} \right.
\label{eqVeff}
\ee
since helicity is either $\pm 1$ near $T_c$.

The resulting masses of bound states are summarized in
Fig.~\ref{fig2}. As shown in the figure, the mass of bound states
approaches zero as $T$ goes to $T_c$, independently of the thermal
quark masses. The factor of 2 introduced by the magnetic
interactions is precisely the factor needed to make the (zero)
masses of $\pi$ and $\sigma$ continuous across $T_c$, as dictated
by chiral symmetry (since the pion mass is protected and the
$\sigma$ is degenerate with the pion at $T_c$).

\section{Equilibrium Above $T_c$ and STAR $\rho^0/\pi^-$ Ratio}

In a recent experiment, STAR~\cite{star} has reconstructed the
$\rho$-mesons from the two-pion decay products in the Au + Au
peripheral collisions at $\sqrt{s_{NN}}= 200$ GeV. They find the
ratio of
 \be
\frac{\rho^0}{\pi^-}|_{STAR}=0.169\pm 0.003 ({\rm stat})\pm 0.037
({\rm syst}),\label{stardata}
 \ee
almost as large as the $\rho^0/\pi^-=0.183\pm 0.001 ({\rm
stat})\pm 0.027 ({\rm syst})$ in proton-proton scattering.
If one assumes equilibrium at the freezeout, then the
ratio is expected to come to about $\frac{\rho^0}{\pi^-}\sim
4\times 10^{-4}$~\cite{peteretal}.

In describing what happens as the system
expands and the temperature cools from $T_c$ down to $T_{flash}$,
we choose the 32 degrees of freedom for the $\rho$, $\pi$,
$\sigma$ and $a_1$ in the $SU(4)$ multiplet (for up and down
flavors) that come down from above $T_c$ as described in
\cite{BLR05,BGR}. These are the light degrees of freedom found in
the quenched lattice calculation of Asakawa et
al.~\cite{asakawaetal} and used in \cite{BLRS}. We suggest that
the entropy at $T_c$ in lattice calculations correspond to that of
massless bosons. Below $T_c$, the hadronic freedom is operative
with the mesons nearly massless (due to the chiral symmetry restoration
at $T_c$).
Now in Harada-Yamawaki HLS/VM \cite{HY:PR}, near $T_c\approx 175$
MeV, the width drops rapidly as the mass drops:
 \be
\frac{\Gamma^\star_\rho}{\Gamma_\rho} \sim
\left(\frac{m_\rho^\star}{m_\rho}\right)^3
\left(\frac{g^\star}{g}\right)^2 \Rightarrow
\left(\frac{m_\rho^\star}{m_\rho}\right)^5.
 \ee
We assume that the effective gauge coupling in medium denoted
$g^*$ begins to scale only above $T_{flash}\approx
T_{freezeout}\approx 120$ MeV in peripheral collisions when the
soft glue has begun to melt. This is analogous to the behavior in
density where it is empirically established that the scaling of
$g^*/g$ sets in only at $n\sim n_0$~\cite{BR2004}.
With nearly zero couplings and masses, the particles stream
freely without interaction until the vector mesons go about
90\% on-shell at the flash temperature $T_{flash}\approx 120$ MeV
at which the soft glues condensate and induce strong interactions
triggering the decay into pions.

We find that in total 63 pions result at the end of the first
generation from the 32 $SU(4)$ multiplet, i.e., $\rho$ (18), $a_1$
(27), $a_0$ (4), $\pi$ (3), $\sigma$ (2) and $\epsilon\equiv f
(1285)$ (12) where the number in the parenthesis is the number of
pions emitted \cite{BLR:STAR}. Excluded from the counting are the $\omega$ and
$\eta$ since they leave the system before decaying. Also left out
are the three $\pi^-$'s coming from the $\rho^0$ decays which are
reconstructed in the measurement. One-third of the pions counted
will be $\pi^-$'s, so there will be 21 $\pi^-$'s. The original 3
$\rho^0$ can be reconstructed in the STAR projection chamber, so
we find \cite{BLR:STAR}
  \be
\frac{\rho^0}{\pi^-}\approx 3/21\approx 0.14, \label{prediction}
 \ee
which is in good agreement with the observed value (\ref{stardata})
considering that the standard scenario would be off by several
orders of magnitude \cite{peteretal}.

The factor of $> 400$ enhancement with respect to the equilibrium
value can be understood as follows. Because of the decreased width
due to the hadronic freedom, the $\rho$ meson goes through {\it
only one generation} before it freezes out in the peripheral
collisions in STAR. But there is no equilibrium at the end of the
first generation. Had the $\rho$ possessed its on-shell mass and
width with $\sim$ five generations as in the standard scenario, it
is clear that the ${\rho^0}/{\pi^-}$ ratio would be much
closer to the equilibrium value.

\section{Conclusions}

By analyzing the Bielefeld lattice gauge results, we found that the
masses of $\pi$ and $\sigma$ approaches zero at $T_c$ from above,
the resulting zero masses are continuous across $T_c$, as dictated by
chiral symmetry (since the pion mass is protected and the $\sigma$ is
degenerate with the pion at $T_c$).
We also show that the surprisingly large STAR ratio can be simply
understood in terms of the (massless) hadronic degrees of freedom above $T_c$.

\end{document}